\documentclass[twocolumn]{aastex61}
\newcounter{ion}

\shorttitle{An X-ray Study of Two B+B Binaries}
\shortauthors{Ignace et al.}

\begin{document}
\title{An X-ray Study of 
Two B+B Binaries:  AH~Cep and CW~Cep}

%% Use \author, \affil, and the \and command to format
%% author and affiliation information.
%% Note that \email has replaced the old \authoremail command
%% from AASTeX v4.0. You can use \email to mark an email address
%% anywhere in the paper, not just in the front matter.
%% As in the title, use \\ to force line breaks.

\author{R. Ignace\email{ignace@etsu.edu}}
\affil{Department of Physics and Astronomy, East Tennessee State 
University, Johnson City, TN 37663, USA }
\author{K.T. Hole}
\affil{Department of Physics, Norwich University, Northfield, VT 05663 USA}
\author{L.M. Oskinova}
\affil{Institute for Physics and Astronomy, University Potsdam, 
14476 Potsdam, Germany}
\affil{4 Kazan Federal University, Kremlevskaya Str., 18, Kazan, Russia}
\author{J.P. Rotter}
\affil{Department of Physics, Norwich University, Northfield, VT 05663 USA}

%% Notice that each of these authors has alternate affiliations, which
%% are identified by the \altaffilmark after each name.  Specify alternate
%% affiliation information with \altaffiltext, with one command per each
%% affiliation.

%\altaffiltext{1}{Visiting Astronomer, Cerro Tololo Inter-American Observatory.
%CTIO is operated by AURA, Inc.\ under contract to the National Science
%Foundation.}
%\altaffiltext{2}{Society of Fellows, Harvard University.}
%\altaffiltext{3}{present address: Center for Astrophysics,
%    60 Garden Street, Cambridge, MA 02138}
%\altaffiltext{4}{Visiting Programmer, Space Telescope Science Institute}
%\altaffiltext{5}{Patron, Alonso's Bar and Grill}

\begin{abstract}    

AH Cep and CW Cep are both early B-type binaries with short orbital
periods of 1.8~d and 2.7~d, respectively.  All four components are B0.5V
types.  The binaries are also double-lined spectroscopic and eclipsing.
Consequently, solutions for orbital and stellar parameters make the
pair of binaries ideal targets for a study of the colliding winds
between two B~stars.  {\em Chandra} ACIS-I observations were obtained
to determine X-ray luminosities.  AH~Cep was detected with an unabsorbed
X-ray luminosity at a 90\% confidence interval of $(9-33)\times 10^{30}$
erg s$^{-1}$, or $(0.5-1.7)\times 10^{-7} L_{\rm Bol}$, relative to the
combined Bolometric luminosities of the two components.  While formally
consistent with expectations for embedded wind shocks, or binary wind
collision, the near-twin system of CW~Cep was a surprising non-detection.
For CW~Cep, an upper limit was determined with $L_X/L_{\rm Bol} <
10^{-8}$, again for the combined components.  One difference between
these two systems is that AH~Cep is part of a multiple system.  The X-rays
from AH~Cep may not arise from standard wind shocks nor wind collision,
but perhaps instead from magnetism in any one of the four components of
the system.  The possibility could be tested by searching for cyclic X-ray
variability in AH~Cep on the short orbital period of the inner B~stars.

\end{abstract}

%% Keywords should appear after the \end{abstract} command. The uncommented
%% example has been keyed in ApJ style. See the instructions to authors
%% for the journal to which you are submitting your paper to determine
%% what keyword punctuation is appropriate.

\keywords{
    Stars: early-type
--- Stars: individual: AH Cep
--- Stars: individual: CW Cep
--- Stars: massive
--- Stars: winds, outflows 
--- X-rays: binaries}

\section{Introduction}

The modern era has brought forth a plethora of intriguing results
for the study of massive stars based on X-ray observations
\citep[e.g.,][]{2016AdSpR..58..739O}.  Massive star binaries with
colliding winds (colliding wind binaries, hereafter ``CWBs'')
have been a staple of X-ray studies, both theoretically and
observationally \citep[e.g.,][]{1992ApJ...389..635U,1992ApJ...386..265S,
2016AdSpR..58..761R}.  The attractions for stellar astronomers have
been the prospects of luminous and hard X-ray emissions from CWBs,
combined with possibilities for inferring or constraining wind
properties (such as mass-loss rates $\dot{M}$), orbital properties,
and interesting plasma physics (instabilities, possibly magnetism, or
non-equilibrium effects).

\begin{table*}
\begin{center}
\caption{Stellar Parameters\label{tab1}}
\begin{tabular}{lcccc}
\hline\hline & AH Cep & AH Cep & CW Cep & CW Cep \\ \hline
 & (primary) & (secondary) & (primary) & (secondary) \\ 
Sp.\ Type & B0.5n & B0.5n & B0.5 & B0.5 \\ 
$T_{\rm eff}$ (K)  & 30,000 & 29,000 & 28,000 & 28,000 \\ 
Mass ($M_\odot$)  & 15.4 & 13.6 & 13.5 & 12.1 \\ 
Radius ($R_\odot$)  & 6.4 & 5.9 & 5.7 & 5.2 \\ 
Luminosity ($10^4\, L_\odot$)  & 2.9 & 2.1 & 1.9 & 1.4 \\ 
$v_{\rm esc}$ (km s$^{-1}$)  & 960 & 940 & 950 & 940 \\ 
$v_\infty^\dag$ (km s$^{-1}$) & 1400 & 1400 & 1400 & 1400 \\ 
$\dot{M^\dag}$ ($10^{-9}\,M_\odot$ yr$^{-1}$)  & 2.5 & 2.2 & 1.8 & 1.8 \\ \hline
\end{tabular}

{\small $^\dag$ The wind terminal speeds and mass-loss rates are not measured
properties, but estimated ones.  See Sect.~\ref{sec:pred}.}

\end{center}
\end{table*}

Advances in the modern era have been driven by increases in the
numbers of objects that have been studied via surveys
\citep[e.g.,][]{2011ApJS..194....7N}, plus intensive studies of a
limited number of especially interesting targets \citep[some
recent examples include][]{2015A&A...573A..43L, 2016A&A...590A.113G,
2017ApJ...838...45C}.  One omission to the effort has been the
neglect of CWBs consisting of B+B stars.  Much of the previous focus
on CWBs has involved systems in which one component is a Wolf-Rayet
(WR) star.  The reason is clear:  WR~stars have fast winds and
generally large mass-loss rates that can result in strong X-ray
emissions \citep{1976SvA....20....2P,1976SvAL....2..138C}.  With
modern X-ray telescopes, interest has also been shown in what are
usually  X-ray fainter O+O binaries
\citep[e.g.,][]{2010MNRAS.403.1657P,2016A&A...589A.121R}.  Owing to
low mass-loss rates, B+B
binaries have not, to our knowledge,
been modeled in hydrodynamic simulations, nor
the subject of a dedicated observational study.

\cite{2007A&A...467..647Y} reported on a study of the B+B binary
CV~Vel.  In that paper the authors summarized the properties for
17 fairly short-period and mostly main sequence B+B binaries.  What
makes this list special is that all of the systems are both
double-lined spectroscopic and eclipsing systems with relatively
short orbital periods.  Analyses from a
variety of authors have provided for well-constrained orbital and
stellar parameters of these systems, including masses, radii,
temperatures, luminosities, semi-major axes, and eccentrities, among
other things.  In particular, the viewing inclinations are known
to be near edge-on.

As a result, we selected the two most massive binaries of the listing
-- AH Cep and CW Cep -- with the intent of detecting evidence for
a wind-wind collision between B stars using the {\em Chandra}
X-ray Telescope.  This paper describes
expectations for the observations, and reports on the curious result
of one detection and one non-detection, despite the two systems 
being near twins in their physical parameters.
Section~\ref{sec:pred} describes
target selection and predicted X-ray levels.  Section~3 details the
observations.  Results from the pointings are discussed in relation
to these expectations in Section~\ref{sec:disc}, with concluding
remarks offered in Section~\ref{sec:conc}.

\section{Predicted X-ray Emissions for CWBs}		\label{sec:pred}

\subsection{Target Selection}	\label{subsec:targ}

Table~8 of \cite{2007A&A...467..647Y} lists 17 B+B binaries along with
primarily stellar properties of the components, plus the orbital
period ($P_{\rm orb}$).  All but one of the binaries have orbital
periods under 1 week.  The binary pairs usually consist of the same
spectral type, from B0.5V+B0.5V to B9.5V+B9.5V, although a few
systems consist of pairs with different spectral-type components
(one is B9V+A0V).

The two most massive binaries are comprised of B0.5~V stars:  AH~Cep
(B0.5Vn+B0.5Vn) and CW~Cep (B0.5V+B0.5V).  Table~\ref{tab1} summarizes
the stellar properties of these systems; Table~\ref{tab2} summarizes
their orbital properties.  We adopt the standard usage that the primary
is the more massive star, and the secondary is the less massive star,
signified with subscripts ``1'' for primary and ``2'' for secondary.
Note that the mass-loss rates and terminal speeds are not measured
but calculated from models.  This will be discussed further in
Section~\ref{subsec:xrays}.

Although all four stars in these two binaries share the same spectral type
(aside from the peculiar designation ``n''), the stellar properties are
not exactly identical.  Both binaries have mass ratios $q=M_2/M_1$ of about 
0.9. The masses range from 12 to 15 $M_\odot$.
The luminosities range by a factor of 2 from the least luminous
(secondary for CW~Cep) to the most luminous (primary for AH~Cep).
Regarding the orbits, both binaries have orbital periods of $\sim
2$~days, and the orbits are circular \citep[for AH~Cep and CW~Cep,
resp.]{2005AJ....129..990K,2007A&A...467..647Y}.

\begin{figure}
\centering
\includegraphics[width=1.0\columnwidth]{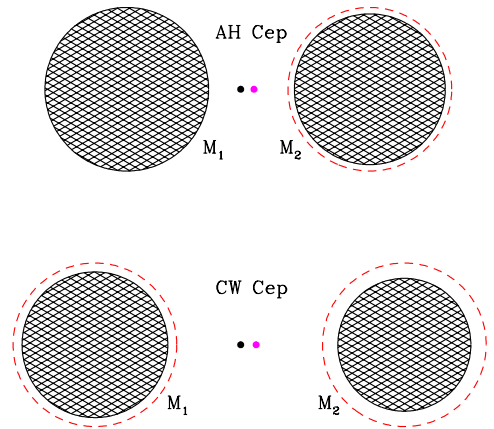}
\caption{A scale representation
of the two binary systems.  Upper
is AH~Cep; lower shows CW~Cep.  Primary and secondary stars are
indicated by the labels $M_1$ and $M_2$, respectively.  The dashed red circles
represent the size of the primary for AH~Cep, as a reference guide.
The solid black dot is the center
of mass.  The stars are situated in the figure such that the center
of mass is at the origin of the coordinate system for each binary.
The magenta dot signifies our estimate for the location of the
colliding wind stagnation point (see Sect.~\ref{sec:pred}).
}
\label{fig1} 
\end{figure}

Figure~\ref{fig1} provides a schematic for the relative sizes and
separations of the stars.  The two systems are displayed on the
same scale and offset vertically from one another.  Between the
stars, the black dot indicates the location of the center of mass
(CM).  The figure is arranged so that the respective CM~points are
the coordinate origins for the two systems.  Nearby in magenta
are the respective stagnation points, as discussed further in
Section~\ref{subsec:xrays}.  The primary for AH~Cep is the largest of
the four stars, and its size is shown as a red dashed circle around the
other three stars for reference.

Ultimately, major considerations for the selection of AH Cep and
CW Cep included:

\begin{enumerate}

\item The two binaries are the most massive ones in the list of
\cite{2007A&A...467..647Y}, suggesting they will be the most X-ray
luminous, even without detection of a colliding wind interaction
(hereafter, CWI), based on the scaling that $L_X \sim 10^{-7} L_{\rm
Bol}$ for massive, single stars
\citep[e.g.,][]{1997A&A...322..167B,2009A&A...506.1055N}.

\item For each binary the component stars are nearly identical.
This suggests that the winds will be nearly identical as well, and
so the CWI will be located close to the CM for each of the respective
systems.  This should simplify the intrepretation of any detected emissions
from the CWIs.

\item Given that the stars are so similar in mass, size, and binary
orbit (inclination and period), observed differences in X-rays could
more confidently be related to ``secondary'' considerations, such
as stellar magnetism.

\end{enumerate}

\subsection{Expected X-ray Properties}	\label{subsec:xrays}

The driving goal for obtaining {\em Chandra} data for AH~Cep and CW~Cep is
to detect for the first time a ``classical'' wind collision (i.e.,
not involving magnetospheric effects) among B+B binaries.  Failing
in that, we expected to detect X-ray emissions at the level predicted
for single massive stars.

For X-ray emissions from the individual stellar winds, we had
anticipated that, being of quite early types in the B spectral class
at B0.5, the individual stars would follow the relation of $L_X
\sim 10^{-7} L_{\rm Bol}$ \citep[see ][and references
therein]{2016AdSpR..58..739O}.  Table~\ref{tab3} provides estimates
for the X-ray luminosities $L_{X\ast}$, with
subscript ``$\ast$'' referring 
to the stellar components of the binaries.  These values are totals
for the two components, treating each one as adhering to the relation
for single stars.  This canonical scaling for single stars is known
to have significant dispersion, and its extension much into the B
spectral class is recognized as dubious
\citep[e.g.,][]{1997ApJ...487..867C,2009A&A...506.1055N,
2013MNRAS.429.3379O}.

For X-rays from a colliding wind, the situation is far more
speculative.  First, such estimates require knowledge of the mass-loss
rates $\dot{M}$ and terminal speeds $v_\infty$ of the stellar winds.  

\begin{table}[b]
\begin{center}
\caption{Orbital Parameters\label{tab2}}
\begin{tabular}{lcc}
\hline\hline & AH Cep & CW Cep\\ \hline
$P_{\rm orb}$ (d) & 1.78 & 2.73 \\ 
$a$ ($R_\odot$)  & 19.0 & 24.2 \\ 
$e$ & 0.0 & 0.0 \\ 
$q=M_2/M_1$ & 0.88 & 0.90 \\ \hline
\end{tabular}
\end{center}
\end{table}

For a rough estimate of the wind speed, we adopted a scaling from
the theory of \citet[][CAK]{1975ApJ...195..157C} for line-driven
winds.  The first is that the wind terminal speed scales as $v_\infty
\propto v_{\rm esc}$.  For B~stars, 
we adopt a value of 1.5 for the
ratio, for which all four stars are estimated to have $v_\infty\approx 1400$
km~s$^{-1}$, as indicated in Table~\ref{tab1}.

For the mass-loss rates, the following relation was used from
\cite{2000A&A...362..295V}:

\begin{equation}
\log \dot{M} = -22.7 +8.96\, T({\rm kK}) -1.42\, T^2({\rm kK}),
	\label{eq:mdot}
\end{equation}

\noindent which is claimed to have validity
for $T > 15$~kK.  Here the mass-loss rate
is in solar masses per year.  For AH~Cep and CW~Cep, values for
$\dot{M}$ are given in Table~\ref{tab1}.  
In an analysis
of {\em IUE} spectral data, \cite{1996Obs...116...89P} set upper
limits to the $\dot{M}$ values from equation~(\ref{eq:mdot})
for the stars in CW Cep at $1.0\times
10^{-8}~M_\odot$ yr$^{-1}$ and $0.3\times 10^{-8}~M_\odot$ yr$^{-1}$
for the primary and secondary components, respectively.
In the case of CW~Cep, the
values are below the upper limits of \cite{1996Obs...116...89P}.
Given the similarity of the stars in AH~Cep to those of CW~Cep, it is
notable that the $\dot{M}$ values for components of AH~Cep are
also below the upper limits established for CW~Cep.

To obtain estimates for the X-rays from colliding winds, an important
parameter is the ratio of wind momentum rates for the two stars
involved.  This parameter is given by

\begin{equation}
\eta = \frac{\dot{M}_2 v_2}{\dot{M}_1 v_1}.
\end{equation}

\noindent Note that different authors use different symbols and
definitions for the ratio of wind momenta.  The above follows
\cite{2009ApJ...703...89G}.  (By contrast, \cite{2016AdSpR..58..761R}
define $\eta$ as the inverse of the above.)  Instead of using
estimated values of mass-loss rates and terminal speeds, we note
that CAK theory gives $\dot{M}v_\infty \propto M_\ast^{3/2}$, for
which case $\eta \approx q^{3/2}$, assuming that the terminal wind
speeds for the components in each binary are equal.  This further
assumes the shock forms after the respective winds have achieved
terminal speeds, which we signify as $\eta_\infty$.  Values of
$\eta_\infty$ derived in this way are listed in Table~\ref{tab3}.

The $\eta$ parameter determines the location of the stagnation
point, and for an adiabatic wind, it can be used to estimate the
X-ray luminosity \citep{1992ApJ...386..265S} and the opening angle
of the bow shock \citep{2009ApJ...703...89G}.  
Let $x$ be the fractional distance between the stars for the
location of the stagnation point, from the primary.  Then $1-x$ is
the fractional distance from the secondary to that point.
\cite{1992ApJ...386..265S} showed that

\begin{equation}
\zeta = \frac{1-x}{x} = \eta^{1/2}.
\end{equation}

\noindent Consequently, for CAK theory, $\zeta_\infty \approx q^{3/4}$,
again for winds at terminal speed.
Values of $\zeta_\infty$ are given in Table~\ref{tab3}.  The stagnation
point, using this relation, is indicated in the schematic of
Figure~\ref{fig1} by the magenta dot.

\cite{2009ApJ...703...89G} derived the opening angle, $\theta_{\rm
Sh}$ for the bowshock when the cooling is strictly adiabatic.  This
too relates to $\eta$ via the expression $\eta = \tan^4 (\theta_{\rm
Sh}/2)$.  Again, for CAK theory it can be shown that

\begin{equation}
\theta_{\rm Sh} \approx 2 \tan^{-1} \, q^{3/8}.
\end{equation}

\noindent However, for a colliding wind shock that is radiative, 
\cite{1996ApJ...469..729C} determine the opening angle to be
(using the modified version from \cite{2009ApJ...703...89G}):

\begin{equation}
\eta = \frac{\tan \theta_{\rm Sh} - \theta_{\rm Sh}}
	{\tan \theta_{\rm Sh} - \theta_{\rm Sh} + \pi}.
\end{equation}

\noindent Whether adiabatic or radiative, the derived opening angles,
given in Table~\ref{tab3}, are very close to $90^\circ$.  This
indicates that, neglecting the effects of the orbital motion, the
discontinuity surface for the colliding winds should be nearly
planar at the scale of the binary separation.

One challenge to these conclusions is the fact that the stellar
components are so close to one another that the stagnation point
actually lies within the wind acceleration zones of the two stars
in each binary.  Using a standard $\beta=1$ wind velocity law for
illustration, as often invoked for early-type winds, we have that

\begin{equation}
v(r) = v_\infty\,\left(1-\frac{bR_\ast}{r}\right),
	\label{eq:vlaw}
\end{equation}

\noindent where $v$ is the radial velocity of the spherical wind
at radius $r$, and $b \lesssim 1$ sets the speed at the wind base.
For AH~Cep, the wind speed is just over a third of the terminal
value at the predicted stagnation point.  For CW~Cep, the speeds
are just over half of terminal.  In both cases the fractional speeds,
$v/v_\infty$ are almost exactly the same for the respective components,
and so $\eta$ is little changed.  However, because the winds have
not achieved terminal speed, the structure of the CWI is probably
not well-represented by the scenario in which both stars have
achieved terminal speed.  Nonetheless, one may still expect that
the shock discontinuity surface is largely planar between the stars
(again, neglecting Coriolis effects that act to distort the
surface from planar).

\cite{1992ApJ...386..265S} provide relations for the peak temperature
achieved in the wind collision, for whether the cooling is predominantly
adiabatic or radiative, and if the former, a scaling relation for
the X-ray luminosity.  First, the peak temperature can be estimated
as

\begin{equation}
T_{\rm peak} = 13.6~{\rm MK}\,\left(\frac{v}{1000}\right)^2 
	= 1.17~{\rm keV}\,\left(\frac{v}{1000}\right)^2 ,
\end{equation}

\noindent for $v$ in km s$^{-1}$. Again using a $\beta=1$ velocity
law, expected peak temperatures are given in
Table~\ref{tab3}.  Note that the peak value is somewhat soft at
0.3~keV in the case of AH~Cep, but fairly hard at 0.7~keV for CW~Cep.

\cite{1992ApJ...386..265S} provide a relation for the ratio of the
radiative cooling time to the flow time as a discriminant between
predominantly radiative or adiabatic cooling.  The ratio is

\begin{equation}
\chi = t_{\rm cool} / t_{\rm flow} \sim \frac{(v/1000)^4\,(d/10^7)}
	{\dot{M}/10^{-7}},
	\label{eq:chi}
\end{equation}

\noindent with $v$ in km s$^{-1}$, $d$ the instantaneous
binary separation in km, and $\dot{M}$ the mass-loss rate.
in $M_\odot$~yr$^{-1}$.  The value of $\chi$ is about
2 for AH~Cep and 20 for CW~Cep, with $\chi \gtrsim 1$ indicating
that the cooling is adiabatic.

With adiabatic cooling the X-ray luminosity is estimated with

\begin{equation}
L_X \propto \dot{M}^2\,v^{-3.2}\,d^{-1}\,(\eta^2+\eta^{3/2}).
	\label{eq:adiabatic}
\end{equation}

\noindent This expression is a proportionality.  To calibrate, we
use model cwb2 from \cite{2010MNRAS.403.1657P} for a wind collision
between identical stars of type O6~V.  That model, characterized by
$\chi=19$, predicts an X-ray luminosity of $L_X = 1.6\times 10^{33}$
erg~s$^{-1}$ in the 0.5-10 keV band.  From this model, along with
the above equation, X-ray luminosities
can be estimated,
and these are provided as {\em expected} values for the target
sources in Table~\ref{tab3}.  A reminder that subscript ``$\ast$''
refers to stellar components, and ``CWI'' refers to contributions
from the colliding wind interaction.  Note that the predicted
values for $L_{X,CWI}$ 
given in Table~\ref{tab3} are lower limits from evaluation
at wind terminal speed; if the CWI forms at a lower wind speed, the
expected values would be higher, based on 
equation~(\ref{eq:adiabatic})\footnote{Note that in the acceleration
zone of the winds, peak temperature in the post-shock gas may no
longer occur along the line of centers for the stars, but in an
annulus about that line, owing to the condition of oblique shocks, with
possible consequences for the expected X-ray luminosity.  Whether
this can occur requires evaluation via hydrodynamical simulation.}.

\section{Observations with ACIS-I}

\begin{table}
\begin{center}
\caption{Predicted X-ray Properties \label{tab3}}
\begin{tabular}{lcc}
\hline\hline & AH Cep & CW Cep\\ \hline
$d$ (pc)  & 1020 & 640 \\ 
$E(B-V)$ & 0.51 & 0.59 \\ 
$N_{\rm H}$ ($10^{21}$ cm$^2$) & 3.0 & 3.4 \\ 
$L_{\rm X,\ast}$ ($10^{30}$ erg s$^{-1}$)  & 19 & 12 \\ 
%$f_{\rm X,\ast}$ ($10^{-14}$ erg s$^{-1}$ cm$^{-2}$)  & 15 & 24 \\ 
$\eta_\infty$ & 0.83 & 0.85  \\ 
$\zeta_\infty$ & 0.91 & 0.92 \\ 
$\theta_{\rm Sh}$ (radiative) & $87^\circ$ & $88^\circ$ \\ 
$\theta_{\rm Sh}$ (adiabatic) & $87^\circ$ & $87^\circ$ \\ 
$kT_{\rm peak}$ (keV) & 0.29 & 0.69 \\ 
$\chi$ & 2 & 20 \\ 
$L_{\rm X,CWI}$ ($10^{30}$ erg s$^{-1}$)  & $> 4.6^\dag$ & $> 2.0^\dag$ \\ \hline
%$f_{\rm X,CWI}$ ($10^{-14}$ erg s$^{-1}$ cm$^{-2}$)  & $> 3.7^\dag$ & $> 4.1^\dag$ \\ \hline
\end{tabular}

{\small $^\dag$ Values scaled from 
\cite{2010MNRAS.403.1657P}, evaluated at terminal speed.  At
less than terminal speed, eq.~(\ref{eq:adiabatic}) indicates a larger
X-ray luminosity.}

\end{center}
\end{table}

\begin{table}
\begin{center}
\caption{Measured X-ray Properties \label{tab4}}
\begin{tabular}{lcc}
\hline\hline & AH Cep & CW Cep\\ \hline
RA$^a$ & 22 47 52.9414 & 23 04 02.2185 \\
DEC$^a$ & +65 03 43.797 & +63 23 48.718 \\
Exp.\ (ks) & 10 & 7 \\
Counts & 37 & --- \\
$\dot{C}$ (cps) & 0.0037 & $<0.00033$ \\
$HR$ & 0.9 & --- \\
$f_{\rm ACIS}^{b,c}$ ($10^{-14}$ erg s$^{-1}$ cm$^{-2}$)  & 7--24 & $<2$ \\ 
$L_{\rm ACIS}^c$ ($10^{30}$ erg s$^{-1}$)  & 9--33 & $<1.0$ \\ \hline
\end{tabular}

{\small $^a$ RA and DEC taken from GAIA DR1 catalog.}
{\small $^b$ ACIS-I fluxes are the ``unabsorbed'' values.}
{\small $^c$ ACIS-I fluxes and luminosities are for the range
0.3--10~keV.  The ranges quoted for AH~Cep are 90\% confidence
intervals.}

\end{center}
\end{table}

Two observations of the systems were obtained by the Chandra X-ray
Observatory, using the Advanced CCD Imaging Spectrometer (ACIS,
\citealt{1992aiaa.confQ....G}). ACIS-I was chosen for maximum
sensitivity with the ability to perform some spectral analysis with
a relatively short observing time.  Our exposures were designed to
yield similar numbers of X-ray counts, based on the discussion of the
previous section.  The final exposure times for the two observations
were 7~ks and 10~ks for CW~Cep and AH~Cep, respectively. Ephemeris data
from the AAVSO\footnote{https://www.aavso.org/} indicates that neither
system was in eclipse at the time of observation (with confirmation from
\cite{2002AJ....123.2724H} for CW~Cep).

Hydrogen column densities were estimated
using values of $E(B-V)$ from the relation that $N_H \approx 5.8\times
10^{21}$~cm$^2 \times E(B-V)$ \citep{2000asqu.book.....C}.  Observed
colors were combined with expected ones based on spectral class using
\cite{2000asqu.book.....C}. Count rates for ACIS-I were then estimated
with interstellar extinction included.  Note that the thermal plasma
of OB~star X-rays is typified by temperatures of a few MK (i.e., $kT$
values of a couple tenths of a keV \cite[e.g.,][]{2016AdSpR..58..739O}).
Table~\ref{tab4} summarizes information about the observations, such as
the exposure ''Exp'', observed counts, the count rate $\dot{C}$, the 
hardness ratio $HR$, and the inferred fluxes $f$ and luminosities $L$ with
ACIS-I.

\begin{figure}
\centering
\fbox{\includegraphics[width=0.45\columnwidth]{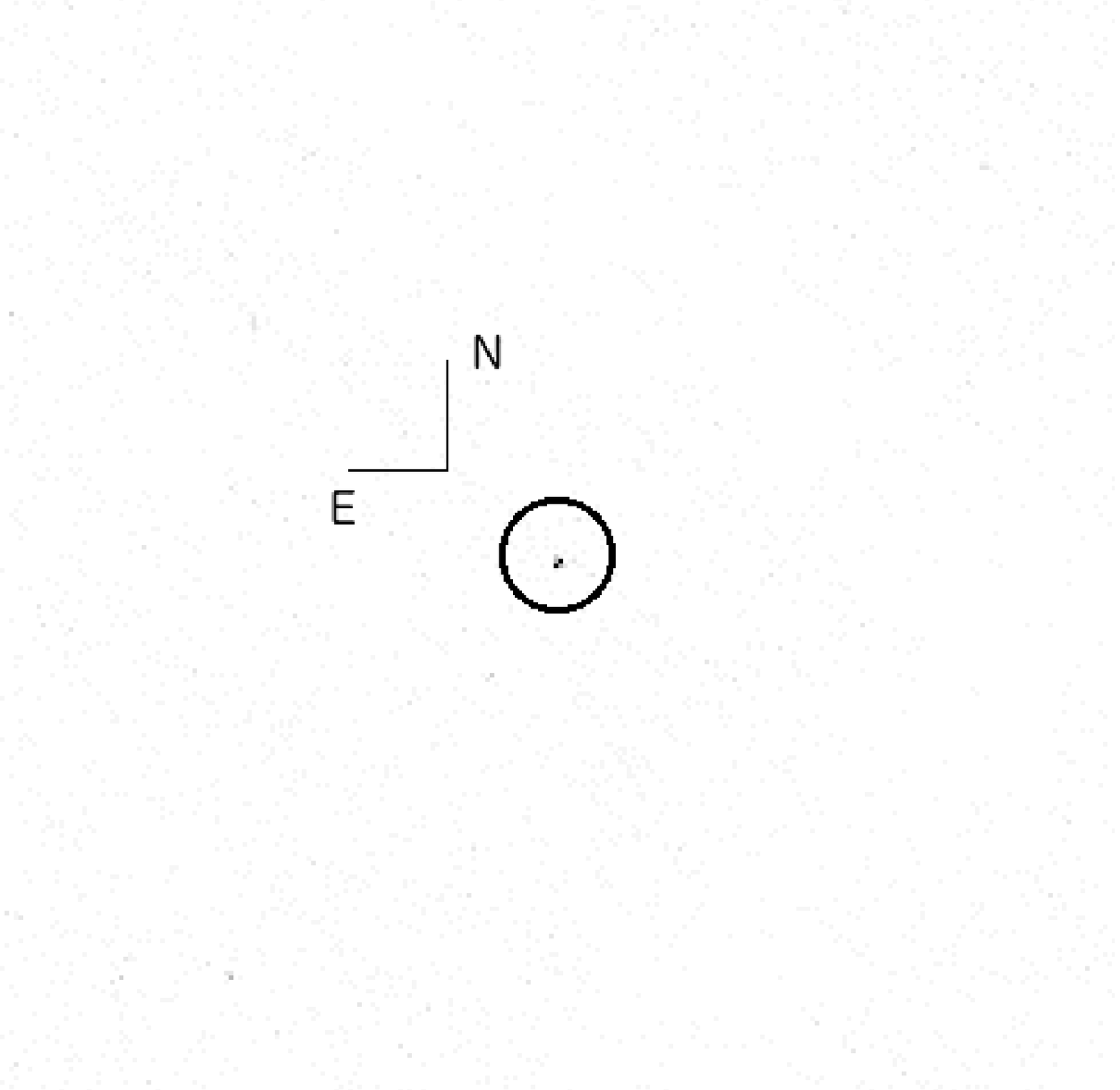}}
\fbox{\includegraphics[width=0.45\columnwidth]{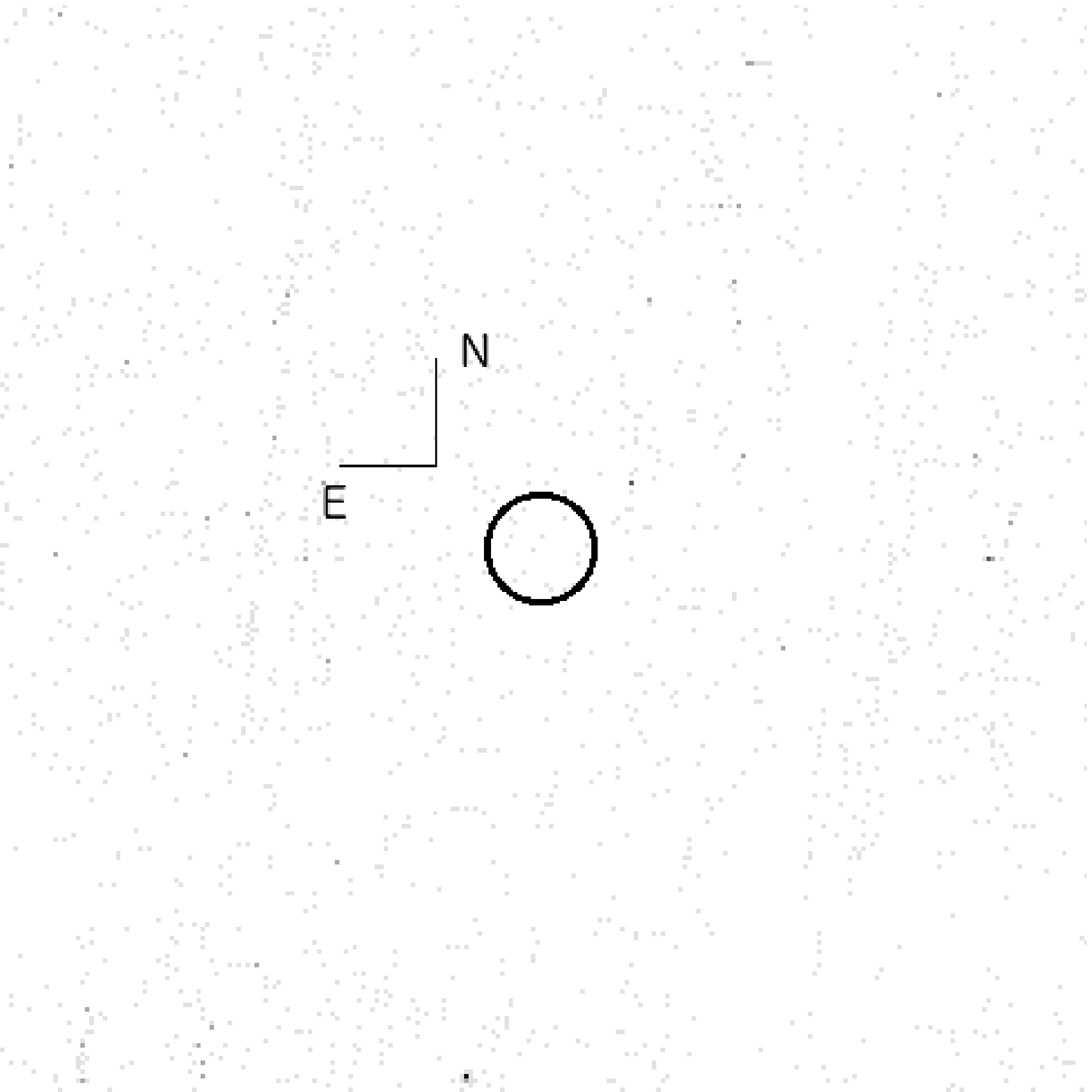}}
\caption{{\em Chandra} ACIS-I image of the field-of-view
for the targets AH~Cep (left; detected) and CW~Cep
(right; undetected).  
Each image is $2'\times 2'$.  The circles of $6~\arcsec$ radius are centered
on source coordinates from SIMBAD.}
\label{fig2} 
\end{figure}

Analysis of both observations was performed with the 
\software{CIAO\footnote{available at http://cxc.harvard.edu/ciao/}
\citep[v4.9; ][]{2006SPIE.6270E..1VF}} software after
standard pipeline processing of the event files.  Source
luminosity was estimated using the srcflux function, assuming
an APEC model, by \citet{1990ARA&A..28..215D} via the HEASARC
database\footnote{https://heasarc.gsfc.nasa.gov/cgi-bin/Tools/w3nh/w3nh.pl}.
Figure~\ref{fig2} displays the field of view for our two targets.
Note that based on the GAIA DR1 Catalog, there are no other objects within
2 arcsec from either target.  The overall 90\% uncertainty circle of {\em
Chandra's} X-ray absolute position has a radius of 0.8~arcsec. The 99\%
limit for the positional accuracy is 1.4 arcsec. The worst case offset
is 2.0 arcsec, but that is for off-axis observations, whereas both of
our pointings were on-axis. For CW Cep, we detected no source photons,
giving us an upper limit on the model luminosity of $\sim 1\times10^{30}$
erg s$^{-1}$. AH Cep was detected with 37 source counts, and an implied
luminosity of $(9-33)\times10^{30}$ erg s$^{-1}$, for a 90\% confidence
interval. Though the S/N is too low to be definitive, AH~Cep emission
does show some energy dependence, and is centered around $\approx 1$ keV.
We used a $1T$ fit with $kT$ of 0.3 keV and 0.6 keV, typical OB star
X-ray spectra, and of the expected temperature for the colliding wind
shock as indicated in Table~\ref{tab3}.

Although the detection of AH~Cep yields an inferred X-ray
luminosity that is basically commensurate with expectations for
embedded wind shocks and/or a wind-collision shock, the non-detection
of CW~Cep, a near twin of the AH~Cep system, makes this interpretation
problematic.

\section{Discussion}	\label{sec:disc}
%\section{Conclusions}

How are we to understand the detection for AH~Cep along with the
non-detection of CW~Cep, given the quite similar properties of these
two early-type binary stars?  Uncertainties in distance and
interstellar absorption could perhaps be important.  However, given
that one of the sources is detected and the other is not, such
effects must conspire to produce an order of magnitude
difference between the two source luminosities.

Alternatively, it is useful to review the assumptions of
Section~\ref{subsec:xrays} for the target sources.  In the theoretical
study of \cite{2013MNRAS.429.3379O} for the scaling between $L_X$ and
$L_{\rm Bol}$ for single massive stars, it seems likely that embedded wind
shocks for early B-type stars, even B0.5~V stars such as our targets,
will likely be adiabatic and therefore faint.  In terms of the emission
expected from the stars individually, we appear to have overstimated
expected X-ray luminosities.

Regarding the colliding wind shock for the respective binary targets,
\cite{2004ApJ...611..434A} derived a convenient expression for the
condition in which the colliding wind shock is radiative or adiabatic
in terms of orbital period.
Their equation~(24) is reproduced here:

\begin{equation}
P < 26~{\rm day}~\left( \frac{20\,M_\odot}{M_1+M_2}\right)^{1/2}\,
	\frac{\dot{M}_{-6}^{1.5}}{v_{1000}^{7.5}},
	\label{eq:orb}
\end{equation}

\noindent where $P$ is the orbital period, $\dot{M}$ is in
$10^{-6}\,M_\odot\,{\rm yr}^{-1}$, and $v$ is the pre-shock speed
relative to 1000~km~s$^{-1}$.  The shock will be radiative when the
inequality is met; otherwise it will be adiabatic.

In the case of AH~Cep, the orbital period is 1.78~days.  With a
terminal speed of 1400~km~s$^{-1}$, the division between radiative
and adiabatic colliding wind shocks is an orbital period of 0.001~days,
or 36 seconds.  The binary period is of course much greater, thus
predicting an adiabatic shock.  However, neither of the winds for
the two components comes anywhere near achieving terminal speed.
The orbital separation is $19.0~R_\odot$.  The primary has a radius
of $6.4~R_\odot$, and the secondary has a radius of $5.9~R_\odot$.
Again using equation~(\ref{eq:vlaw}), the pre-shock wind speed
of the primary for the distance of the stagnation point is
$0.34v_\infty$.  Using $\dot{M}$ estimated for the primary only,
the threshold for a radiative shock increases to 0.7~days.  Although
still too short for the shock to be radiative, it is much closer, being within
a factor of $\sim 2.5$.

Now consider CW~Cep.  The orbital period is longer at 2.73~days.
The orbital separation is somewhat larger, and the stars are both
somewhat smaller than the components of AH~Cep.  Following the steps
in the example of AH~Cep, the colliding wind shock will be adiabatic
for orbital periods longer than about 0.01~days.

In summary, it appears that the embedded wind shocks for both
AH~Cep and CW~Cep are likely adiabatic as consistent with
equation~(\ref{eq:chi}), suggesting that the X-ray emissions
from the individual winds, if they were in isolation, would be
relatively weak.  The colliding wind shocks are likewise adiabatic
and weak sources.  However, there is tremendous sensitivity of this
criterion to the mass-loss rate and pre-shock wind speed, of which
neither is well-constrained for either system.  If the X-rays of
AH~Cep do originate from the colliding wind shock, it would imply
that we have, for the first time, detected X-rays from a B+B wind
collision\footnote{\cite{2017arXiv170304686P} report the detection of
variable X-rays from the B2IV+B2V binary $\rho$~Oph A+B, but attribute
the X-rays to magnetic effects for the fast-rotating, young primary star.
\cite{2015MNRAS.454L...1S} report on X-rays from B+B binary $\epsilon$~Lup
in which both stars are magnetic with interacting magnetospheres.}.

There is an alternative explanation to account for
the detected X-rays.  Several previous
studies suggest that AH~Cep is a multiple system, with four components
in total \citep{1986IBVS.2886....1M,
1989A&A...221...49D,1990PTarO..53..115H,2005AJ....129..990K}.
Component \#3 is assigned a period of 67.6~years in an eccentric
orbit of $e=0.52$.  Component \#4 has a period of 9.6 years, in an
even more eccentric orbit with $e=0.64$.  The two stars have
respective mass estimates of $M_3 \approx 8 M_\odot$ and $M_4 \approx
4 M_\odot$, making them spectral types B2-B3 and B7-B8, respectively
\citep{2005AJ....129..990K}.  The age of the system is estimated
at about 6~Myrs \citep{1990A&A...236..409H}.
At these spectral types, neither the tertiary or quarternary stars are
expected to be X-ray bright, unless perhaps the stars have magnetic fields
\citep{2011MNRAS.416.1456O,2013MNRAS.429..398P,2014ApJS..215...10N}.

Binarity among massive stars is known to be relatively normal
\citep[e.g., a recent short review by][]{2017arXiv170301608S}.  Less
is known about hierarchical systems amongst massive stars.  
The well-known multiplet massive star system Mintaka
($\delta$~Ori, HD~36486) is bright, relatively close,
and well-studied in many wavebands, including extensive observations at X-ray 
wavelengths \citep{2015ApJ...808...88R, 2015ApJ...809..132C,
2015ApJ...809..133N,2015ApJ...809..134P,2015ApJ...809..135S}. The center of the
system is a triple, involving an O9.5 II primary and an early-type
secondary in a fairly tight orbit of period $\approx 5.7$~days.
A more distant third component of perhaps $\approx 8~M_\odot$ follows
an elliptical orbit of nearly 350 years.  Although Mintaka is an
X-ray source, the bulk of the emission is associated with the
embedded shocks for the primary wind, as opposed to a wind collision
with the secondary star's wind, or as arising with the tertiary.
Mintaka is a case in which the X-ray emissions are dominated by
the primary star, but in contrast to our targets, the primary
for Mintaka is an evolved late-type O~star.

$\beta$~Cru is an example of a triple system involving massive
stars that displays a complex X-ray signal \citep{2008MNRAS.386.1855C}.
The primary is B0.5~III, so the same spectral class as the stars
in our binaries, but a giant instead of a main sequence star.  The
secondary is B2~V \citep{1998A&A...329..137A} in an eccentric orbit with a
 period of 5~years.  The age of the system is estimated at around
8-10~Myrs \citep{2008MNRAS.386.1855C}, which is not much greater
than the estimate for AH~Cep.  Interestingly, \cite{2008MNRAS.386.1855C}
report on a pre-Main Sequence (PMS) companion in their X-ray study of
$\beta$~Cru, betrayed through its relatively hard contribution to
the X-ray emission detected from the system.  $\beta$~Cru represents
a case in which the massive primary does not entirely dominate
the X-ray emissions.  Whereas the primary for $\beta$~Cru is evolved,
the primary and secondary stars in AH~Cep are less luminous main sequence
stars.  Consequently, either/both of the other companions
could have a relatively more important contribution to observed X-ray
emissions, if magnetic.

\section{Conclusions}	\label{sec:conc}

There are three main mechanisms to consider for understanding the
detection of X-rays from AH~Cep but not from CW~Cep.

\begin{enumerate}

\item The X-rays detected in AH~Cep may come from a colliding wind
shock that is either not present or not
detected in CW~Cep.  The predictions for whether
the colliding wind shock is radiative or adiabatic are quite
sensitive to the velocity distribution of the stellar winds, and
somewhat sensitive to the mass-loss rate.  Using a $\beta=1$
velocity law, and given the different separations between
the primary and secondary stars in our targets, we may expect
the colliding wind shock of CW~Cep to be $\sim 3\times$ smaller
than for AH~Cep, yet the upper limit to the X-ray luminosity
for CW~Cep is over $10\times$ smaller.

Our adopted terminal speeds may be too high, or too low.  Moreover,
our use of $\beta=1$ for the velocity law could well be in error:  the
radiation from each of the stars may modify the wind driving inbetween
the stars.  It should be noted that \cite{1989MNRAS.241..721P} provides
an analysis of {\em IUE} spectra for a number of main sequence B~stars,
among them some early types.  Values for $\dot{M}$ and $v_\infty$
are determined only for two B0~V stars, and the mass-loss rates are
actually products, $\dot{M}q$, where $q$ is the ionization fraction for
the species used in the P~Cygni line analysis.  Consequently, the values
from \cite{1989MNRAS.241..721P} provide only lower limits to $\dot{M}$,
which for the two B0~V stars are nearly two orders of magnitude lower
than values adopted from \cite{2000A&A...362..295V}.  The terminal
speeds are also lower than what we have adopted.  Using values from
\cite{1989MNRAS.241..721P} would indicate that the colliding wind
shocks for AH~Cep and CW~Cep are strongly adiabatic.  Although it seems
unlikely that we have detected X-rays from the colliding wind shocks,
the wind properties of the stars and of the colliding wind interaction,
being in the wind acceleration zone, are simply too poorly known.

\item It seems unlikely that we have detected wind embedded shocks
from the individuals winds.  The $L_X/L_{\rm Bol}$ ratio for AH~Cep
is low but commensurate with expectations for single stars.  However,
all four of the B0.5~V stars 
are nearly identical.
Consequently, it is difficult to understand why AH~Cep
is detected when CW~Cep is not, if the X-rays arise from wind shocks.
Perhaps the primary or secondary in AH~Cep is magnetic.  Magnetic B~stars
are known to be diverse in the luminosity and hardness level of
their X-rays \cite[e.g.,][]{2011MNRAS.416.1456O,
2013MNRAS.429..398P,2014ApJS..215...10N}.

\item One distinction between our two targets is that AH~Cep has
been reported to be a multiplet system of 4 stars, whereas CW~Cep appears to
be only a binary.  It is possible that either or both of the other
two stars in AH~Cep are responsible for its X-ray emissions.  The
tertiary and quaternary components are thought to be mid and late
B stars, respectively.
Detectable X-rays from embedded wind shocks for either of these
objects seems highly unlikely, given the roughly
$\dot{M}^2$ dependence of X-ray
luminosity for these spectral types \citep{2013MNRAS.429.3379O}.
Wind collision is an equally unlikely explanation:  the large
separation implies low densities and small emission measures.
The ratio $L_X/L_{\rm Bol}$ for the detection of the AH~Cep system
is $\approx 1\times 10^{-7}$.  If the B2-B3 star of the system
were the source of X-rays, the ratio would increase to $\sim 10^{-6}$
for that object; if the even later B8-B9 star is the source, the
ratio would be $\sim 10^{-5}$.

\end{enumerate}

An interesting implication of the first two points -- namely that
X-rays are not detected from the stellar winds or colliding winds --
would further support the emerging picture that the wind properties
of B~stars are poorly known, and that the winds may be quite weak.
Failure to detect X-ray signatures from colliding winds could be a
combination of low $\dot{M}$ values and low-speed flow.  The latter
would result in weaker shocks of lower temperatures at $\sim 1$~MK.
Failure to detect X-rays from the individual winds leads to the same
conclusion of weak winds.  As further evidence in support of B~stars
having weak winds, \cite{2012A&A...537A..37M} mention difficulties with
obtaining wind solutions for $L \lesssim 2\times 10^5 \, L_\odot$.
Their struggle aligns well with the low $\dot{M}$ and $v_\infty$
values obtained by \cite{2011MNRAS.416.1456O} in their study of several
magnetic B~stars.  The situation for the B~stars is complicated by the
fact that some weak-wind B~stars are relatively strong X-ray emitters
\citep{2012ApJ...756L..34H, 2017AAS...22923205D}.

Progress toward understanding the differences in X-ray emissions
between these two systems can be addressed with new observations.
Certainly, better understanding of the stellar winds would be
furthered by performing a detailed analysis of UV~spectra of the
systems, if possible.  For example, \cite{1996Obs...116...89P} were
only able to derive upper limits to the mass-loss rates for CW~Cep.
A deep X-ray exposure of CW~Cep could allow for a detection of its
faint X-rays, or at least place a more stringent upper limit on
its emission.  An X-ray light curve for the detected source, AH~Cep,
would constrain the source of X-rays. If its X-ray
emissions vary in phase with the orbit period of the primary and
secondary, then the X-rays arise from the inner binary of this
multiplet.  Dimming of the X-rays when either of the stars are forefront
(i.e., during an optical eclipse, twice per orbit) would favor a
colliding wind shock as the source of X-rays, as opposed to embedded wind
shocks. Failure to detect variability of either kind could suggest
stellar magnetism among any of the four components of AH~Cep as an
explanation for the X-ray detection. \\

\acknowledgments

The authors gratefully acknowledge an anonymous reviewer for making
several comments that have improved this paper.
Support for this work was provided by the National Aeronautics and
Space Administration through Chandra Award Number G05-16013A issued
by the Chandra X-ray Observatory Center, which is operated by the
Smithsonian Astrophysical Observatory for and on behalf of the National
Aeronautics Space Administration under contract NAS8-03060.  KTH and JPR
would like to acknowledge the Norwich University Office of Undergraduate
Research and the Office of Academic Research for support of this project.
The work of LMO is partially supported by the Russian Government Program
of Competitive Growth of Kazan Federal University.  This research has
made use of the SIMBAD database, operated at CDS, Strasbourg, France.
We acknowledge with thanks the variable star observations from the AAVSO
International Database contributed by observers worldwide and used in this
research.  This research has made use of data and/or software provided by
the High Energy Astrophysics Science Archive Research Center (HEASARC),
which is a service of the Astrophysics Science Division at NASA/GSFC and
the High Energy Astrophysics Division of the Smithsonian Astrophysical
Observatory.  This work has made use of data from the European Space
Agency (ESA) mission {\it Gaia} (\url{https://www.cosmos.esa.int/gaia}),
processed by the {\it Gaia} Data Processing and Analysis Consortium (DPAC,
\url{https://www.cosmos.esa.int/web/gaia/} \url{dpac/consortium}). Funding
for the DPAC has been provided by national institutions, in particular
the institutions participating in the {\it Gaia} Multilateral Agreement.

%\begin{thebibliography}{}
\bibliography{ignace}
%\end{thebibliography}

\end{document}